\documentclass{Interspeech}

\usepackage{graphicx}
\usepackage{booktabs}
\usepackage{multirow}
\usepackage{amssymb}

\interspeechcameraready

\title{The 1st SpeechWellness Challenge: \\
Detecting Suicide Risk Among Adolescents}

{\author{\centerline{Wen Wu$^{*,1,4}$, Ziyun Cui$^{*,2, 1}$, Chang Lei$^{*,3}$, Yinan Duan$^{3}$, Diyang Qu$^{3}$}\\
\centerline{Ji Wu$^{2}$, Bowen Zhou$^{1,2}$, Runsen Chen$^{\dagger,3}$, Chao Zhang$^{\dagger,1,2}$}}

\address{
 \fontsize{11pt}{13pt}\selectfont{$^1$ Shanghai Artificial Intelligence Laboratory, China \\
  $^2$ Department of Electronic Engineering, Tsinghua University, China \\
  $^3$ Vanke School of Public Health, Tsinghua University, China \\
  $^4$ Department of Engineering, University of Cambridge, UK}\thanks{$*$ Equal contribution. $\dagger$ Corresponding author. Supported by Shanghai Artificial Intelligence Laboratory.}}

\email{\fontsize{9pt}{9pt}\selectfont{wuwen@pjlab.org.cn, cui-zy24@mails.tsinghua.edu.cn, cz277@tsinghua.edu.cn}}


\usepackage{comment}

\begin{document}

\maketitle

\begin{abstract}
    The 1st SpeechWellness Challenge (SW1) aims to advance methods for detecting current suicide risk in adolescents using speech analysis techniques. Suicide among adolescents is a critical public health issue globally. Early detection of suicidal tendencies can lead to timely intervention and potentially save lives. Traditional methods of assessment often rely on self-reporting or clinical interviews, which may not always be accessible. The SW1 challenge addresses this gap by exploring speech as a non-invasive and readily available indicator of mental health. We release the SW1 dataset which contains speech recordings from 600 adolescents aged 10-18 years. By focusing on speech generated from natural tasks, the challenge seeks to uncover patterns and markers that correlate with current suicide risk. 
\end{abstract}

\section{Introduction}

Suicide remains a critical global health challenge and is among the leading causes of death among adolescents worldwide~\cite{world2021suicide,shain2016suicide, orri2020mental}. 
Early detection of suicide risk is essential for effective prevention and intervention of potential suicide attempts. 
The diagnosis of suicide risk is challenging, as individuals exhibit different behaviours, making it difficult to obtain a unified clinical representation of a suicidal individual. It also relies heavily on the participant’s ability, willingness, and honesty in communicating their symptoms, emotions, and thoughts~\cite{cummins2015review}.
Clinical interviews and self-report questionnaires are commonly utilised in clinical practice for diagnosis.
However, clinical interviews are resource-intensive, requiring substantial human effort and the expertise of well-trained clinicians. Questionnaires, although more cost-effective as screening tools, are prone to issues such as bias, discrimination, and deliberate disguise~\cite{ganzini2013trust}.
This motivates the study of automatic suicide risk detection~\cite{oh2017classification,cheng2017assessing,su2020machine,roy2020machine,zhang2021automatic,ghosh2022multitask,dhelim2023artificial}.

Speech has been shown as a promising biomarker for detecting mental disorders~\cite{moore2007critical, valstar2016avec,wu2023self} and cognitive diseases~\cite{luz20_interspeech,li2022alzheimer,cui2023transferring}, as well as suicide risk~\cite{cummins2015review, scherer2013investigating,cui24_interspeech}. Speech provides a rich source of both semantic and non-semantic (paralinguistic) information, and can be collected in a cheap, remote, non-invasive and non-intrusive way. 
From the perspective of semantics, suicidal speech tends to have different top words from non-suicidal speech~\cite{belouali2021acoustic}. 
From the perspective of paralinguistics, difference can be observed in spectral properties, source features (\textit{e.g.}, jitter, shimmer), prosodic features (\textit{e.g.}, F0), formant characteristics~\cite{cummins2015review}, along with increased disfluencies, such as more frequent hesitations and speech errors~\cite{stasak2021read}.

In line with Interspeech 2025's conference theme of \textit{Fair and Inclusive Speech Science and Technology}, we launch the 1st SpeechWellness challenge, with the aim of bringing attention to the importance of mental health in speech technology research and bridging the gap between speech technology, psychology, and healthcare.
To the best of our knowledge, this is the first such challenge focused on suicide risk among adolescents. It is hoped that the challenge could motivate novel algorithms and models that can detect subtle cues in speech related to current suicide risk as well as providing evidence that could inform mental health policies and clinical practices.

\section{Challenge Description}
The 1st {SpeechWellness Challenge} (abbreviated as SW1) focuses on advancing techniques for detecting current suicide risk among adolescents through speech analysis. Challenge teams are tasked with developing models that utilise spontaneous and reading speech as digital biomarkers for binary current suicide risk detection (has current suicide risk or not).

The dataset used for the challenge comprises speech recordings from {600 Chinese teenagers aged 10-18 years}\footnote{The challenge data is shared solely for scientific research purposes after carefully anonymisation and is to be used solely for challenge purposes, with strict prohibitions on redistribution or any other use.}. {All participants (speakers) have been anonymised before making this dataset available to the challenge teams}, which provides a unique opportunity to apply and refine advanced speech technologies for public health while preserving the privacy of all participants. Details about the dataset and anonymisation method will be discussed in Section~\ref{sec: data}. Challenge teams are encouraged to leverage a variety of methodologies, including but not limited to signal processing, self-supervised learning foundation models, speech recognition, emotion recognition, and large language models. The goal is to push the boundaries of current technology to create effective tools for early detection and intervention of suicide risk.

\section{The SW1 Challenge Dataset}
\label{sec: data}
\subsection{Data collection}

Data was collected from 47 elementary and middle schools in Yunfu city, southern China, consisting of 600 participants aged 10 to 18 years.
At-risk students were identified based on professional clinical interviews. The control group includes non-risk students. Informed consent was obtained from all participants and their guardians before data collection, with participants informed of their rights to withdraw at any time or decline uncomfortable questions. Interviews were conducted by trained interviewers specialising in related fields such as psychology and psychiatry, under the supervision of psychiatrists.

\begin{figure}[t]
\begin{minipage}[b]{0.49\linewidth}
    \centerline{\includegraphics[width=\linewidth]{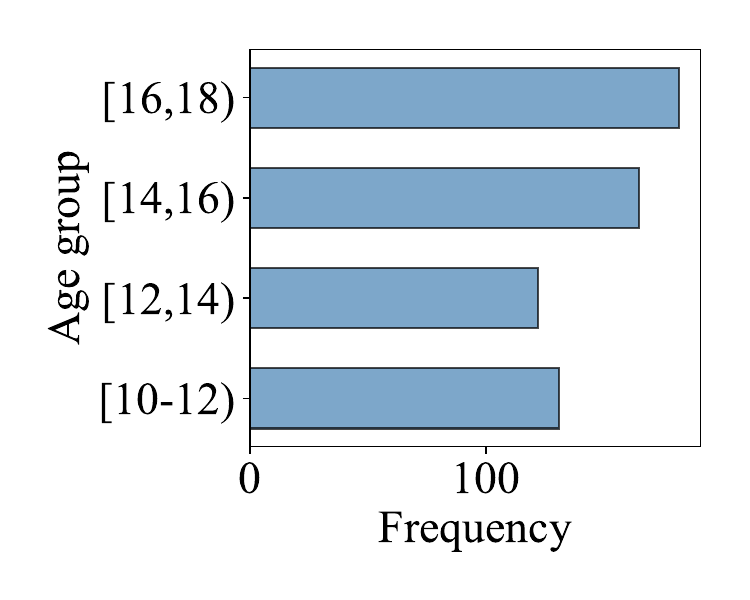}}
    \vskip -1ex
    \caption{Distribution of age group.}
    \label{fig: hist-age}
\end{minipage}
\hfill
\begin{minipage}[b]{0.49\linewidth}
    \centerline{\includegraphics[width=\linewidth]{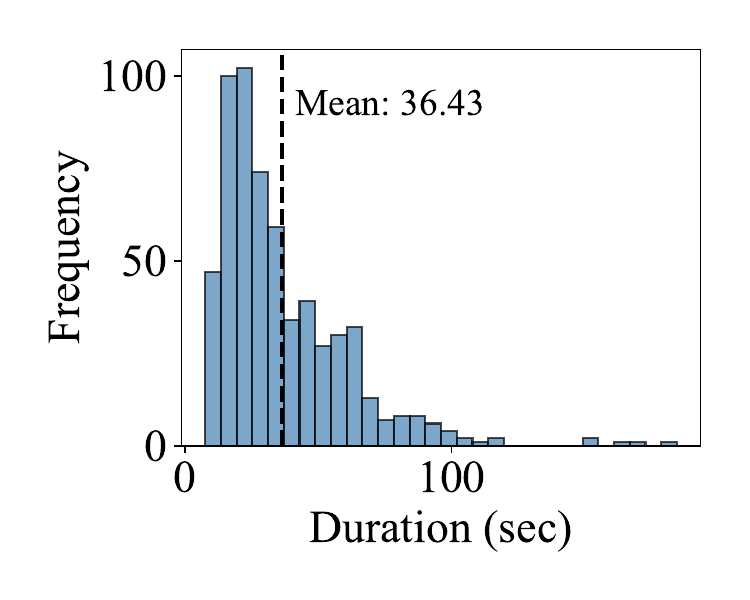}}
    \vskip -1ex
    \caption{Histogram of recording duration of the ER task.}
    \label{fig: hist-ER}
\end{minipage}
\begin{minipage}[b]{0.49\linewidth}
    \centerline{\includegraphics[width=\linewidth]{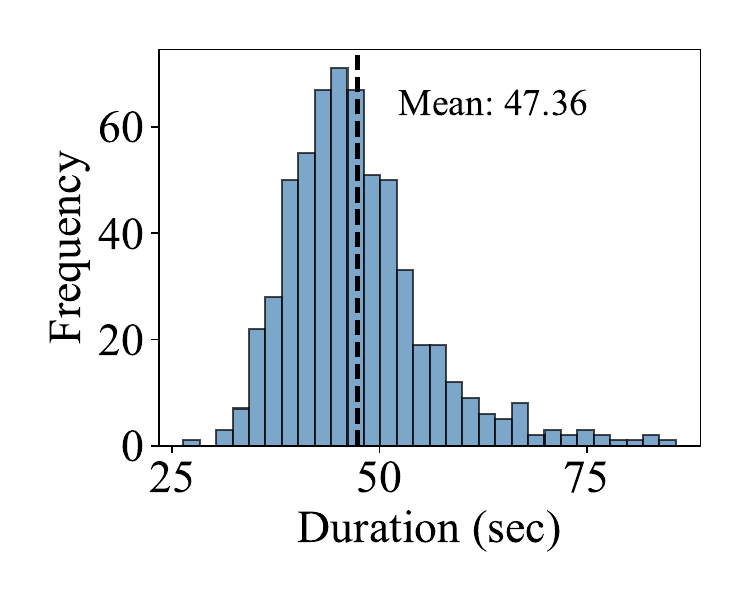}}
    \vskip -1ex
    \caption{Histogram of recording duration of the PR task.}
    \label{fig: hist-PR}
\end{minipage}
\hfill
\begin{minipage}[b]{0.49\linewidth}
    \centerline{\includegraphics[width=\linewidth]{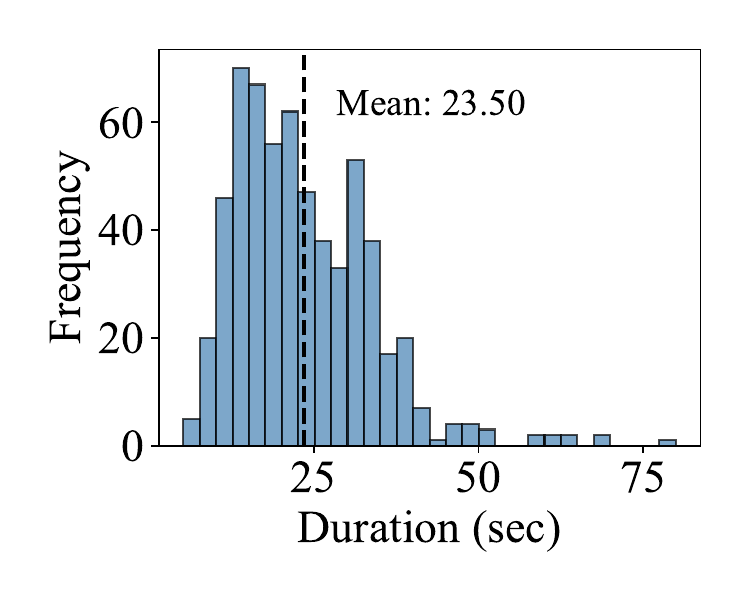}}
    \vskip -1ex
    \caption{Histogram of recording duration of the ED task.}
    \label{fig: hist-ED}
\end{minipage}
\vskip -2ex
\end{figure}

The interview contains both open-ended (unstructured tasks) and structured tasks (such as paragraph reading). Data collection was conducted in quiet, sound-proof rooms with only the interviewer and the participant present to avoid disturbances and maintain privacy. Standardised voice recorders were used by interviewers to capture speech data from participants’ speech tasks, ensuring consistency in audio format and quality. All interviews were conducted in Mandarin. The project has obtained ethical approval from the relevant committee.

The recordings used for the SW1 challenge consist of each participant’s responses to three different speech tasks designed to capture a range of linguistic, cognitive, and emotional features.
\begin{enumerate}
    \item \textbf{Emotional Regulation (ER)}: Participants are asked to answer an open-ended question, ``Have you ever experienced moments of extreme emotional distress? How do you manage such feelings?''
    \item \textbf{Passage Reading (PR)}: Participants read the poetry \textit{The North Wind and The Sun}, a passage from Aesop’s Fables commonly used in linguistic studies for examining structures such as morphemes~\cite{baird2022blowing}\footnote{The passage  can be found at: \url{https://github.com/speechwellness/supplementary-materials/blob/main/README.md}.}. 
    \item \textbf{Expression Description (ED)}: Given an image of a facial expression~\cite{conley2018racially}, participants are ask to describe it\footnote{The image can be found at: \url{https://github.com/speechwellness/supplementary-materials/blob/main/README.md}.}.
\end{enumerate}

Diagnosis of current suicide risk was conducted by professionally trained interviewers based on a professional suicide risk assessment: the Mini International Neuropsychiatric Interview for Children and Adolescents (MINI-KID) suicide diagnostic~\cite{sheehan1998mini}.
It is a structured diagnostic interview designed for adolescents, which is considered a gold standard interview for assessing adolescent psychopathology and has been validated for use by clinicians and researchers to assess current suicide risk based on DSM-IV and ICD-10 criteria~\cite{sheehan2010reliability,liu2011reliability,mcmanimen2020prospective}.

The SW1 dataset comprises speech recordings from 600 participants (420 females and 180 males), evenly divided between 300 non-risk and 300 at-risk individuals. Statistics of age group is shown in Figure~\ref{fig: hist-age}. 
The duration of speech recordings of each task is shown in Figure~\ref{fig: hist-ER}-\ref{fig: hist-ED} with the average duration shown by dashed lines. 
The dataset is split into training, development (dev), and test sets in a 4:1:1 ratio. Consistency in gender distributions across the train/dev/test splits was ensured.
The challenge teams are required to develop a model to predict the label (has current suicide risk or not) for a participant. They are free to choose whether to use recordings from a single task or a combination of all three tasks.

\subsection{The timbre anonymisation process}
Given that the participants are adolescents, a vulnerable population requiring heightened ethical considerations, we adopt additional anonymisation methods to further protect their privacy. Neural voice conversion techniques\footnote{Available at: \url{https://github.com/RVC-Project/Retrieval-based-Voice-Conversion-WebUI}} were applied to alter the timbre of the participants’ voices while preserving prosody and rhythm so that the essential features for developing the automatic suicide risk detection system remain intact. 

\begin{figure}[t]
\centerline{\includegraphics[width=\linewidth]{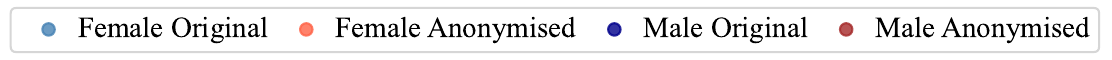}}
\begin{minipage}[b]{0.49\linewidth}
    \centerline{\includegraphics[width=\linewidth]{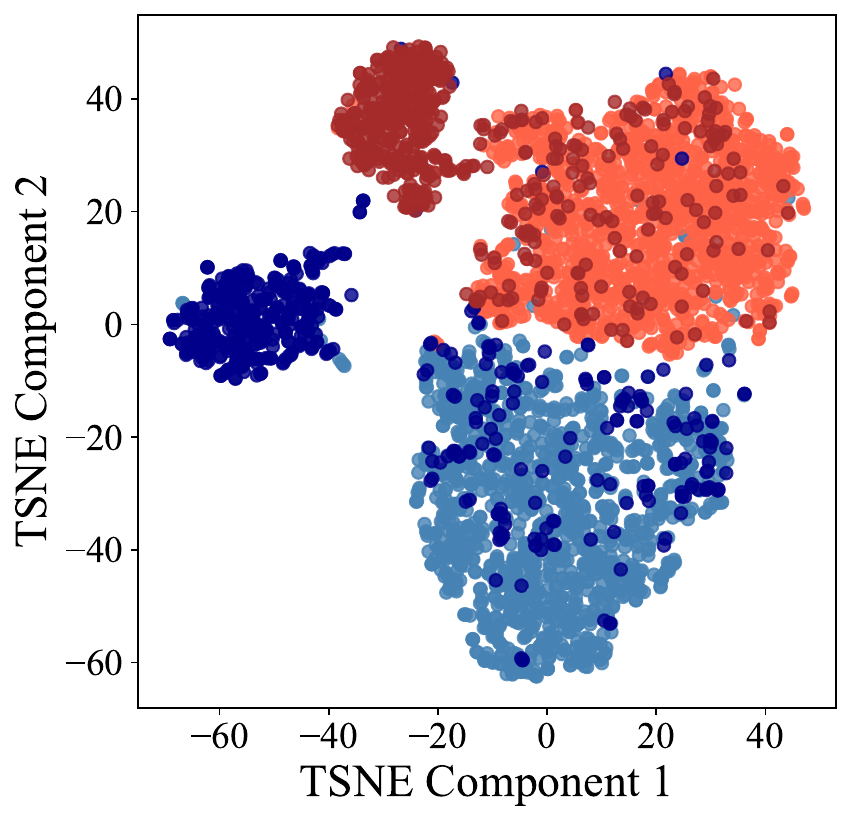}}
    \centerline{(a)}
    \vskip -1ex
\end{minipage}
\hfill
\begin{minipage}[b]{0.49\linewidth}
    \centerline{\includegraphics[width=\linewidth]{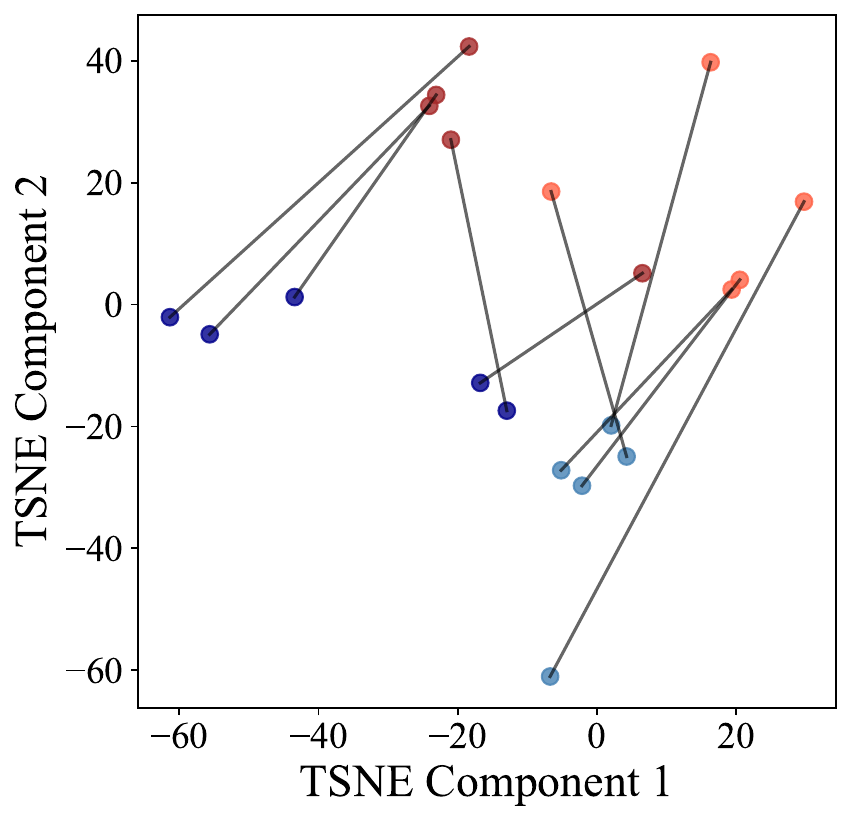}}
    \centerline{(b)}
    \vskip -1ex
\end{minipage}
\caption{(a) TSNE visualisation of speaker embeddings before (blue) and after (red) anonymisation. Dark points represent male speakers, light points represent female speakers. (b) Speaker embeddings of 10 randomly selected cases before and after anonymisation, paired with lines.}
\label{fig:x-vector-2}
\vskip -2ex
\end{figure}

To illustrate the impact of timbre modification, speaker embeddings were extracted from the recordings before and after voice conversion using a pretrained x-vector model\footnote{Available at: \url{https://huggingface.co/microsoft/wavlm-base-plus-sv}}. These embeddings are then visualised in Figure~\ref{fig:x-vector-2}(a). The blue points represent speaker embeddings before anonymisation, while the red points represent embeddings after anonymisation. The figure clearly illustrates that the clusters are distinct and separated. 
The smaller clusters in the upper left, shown in darker colours and set apart from the larger clusters, primarily represent male speakers, while the larger clusters predominantly correspond to female speakers. Since the participants include children in the process of voice maturation, some dark points (male) may appear within the lighter clusters (female).
Figure~\ref{fig:x-vector-2}(b) shows speaker embeddings of 10 randomly selected cases before and after anonymisation, paired with lines to demonstrate the non-linear nature of the anonymisation transformation.

To quantify the impact of anonymisation on speaker verifiability, we compute the equal error rate (EER), which compares pairs of positive (same speaker) and negative (different speakers) utterances based on cosine similarity of x-vectors to verify speaker identity~\cite{tayebi2024addressing}. Post-anonymisation EER is 30.28\% while the original EER is  9.33\%. The increase in EER signifies enhanced anonymisation efficacy.

To further assess the impact of timbre anonymisation on speech intelligibility, the character error rate (CER) was calculated for the original and anonymised recordings\footnote{Recordings of the passage reading task were used, as ground truth transcripts were available.}. We first obtain transcriptions for the original and anonymised recordings using a pretrained ASR system\footnote{Available at: \url{https://www.modelscope.cn/models/iic/speech_paraformer-large-vad-punc_asr_nat-zh-cn-16k-common-vocab8404-pytorch/summary}}, then compute the CER between the ground truth and the transcriptions from the original and anonymised recordings, as well as between the original and anonymised transcriptions. Results are listed in Table~\ref{tab: CER}. The anonymised recordings yield a slightly higher CER than the original recordings, with a 6\% CER between transcriptions before and after anonymisation, which is within an acceptable range.

\begin{table}[t]
    \centering
    \caption{CER between the ground truth and the transcriptions from the original and anonymised recordings, as well as between the original and anonymised transcriptions.}
    \begin{tabular}{cccc}
    \toprule
         \textbf{Ground truth}&  \textbf{Original}&  \textbf{Anonymised}& \textbf{\%CER} \\
         \midrule
         $\surd$&  $\surd$&  & 21.4\\
         $\surd$&  &  $\surd$& 23.8\\
         &  $\surd$&  $\surd$& 6.0\\
         \bottomrule
    \end{tabular}
    \vskip -3ex
    
    \label{tab: CER}
\end{table}

\subsection{Response validation and privacy check}
Apart from the timbre anonymisation process, we also took action to filter out data that doesn't meet the requirement. The validation process consists of four stages:
\begin{enumerate}
    \item Response length: check the response contains sufficient content.
    \item Task relevance: check the interviewee’s responses are relevant to the assigned task.
    \item Sensitive information exposure: ensure that no sensitive privacy-related information is revealed in the response.
    \item Timbre modification quality: check the intelligibility of the speech after voice conversion.
\end{enumerate}
For the response length check, we filtered out recordings with transcriptions shorter than a threshold of 10 words. To assess the speech quality after timbre modification, we set a CER threshold of 40\% between the transcriptions of the original and modified speech and filtered out recordings with a CER exceeding this threshold.
We employed GPT-4o\footnote{\url{https://platform.openai.com/docs/models\#gpt-4o}} to check task relevance and sensitive information exposure. The instructions were presented in a few-shot manner. 
Along with the prompt, we provided a positive and a negative case to help the agent better complete the task\footnote{Detailed prompts can be found at \url{https://github.com/speechwellness/supplementary-materials/blob/main/README.md}.}.

\begin{figure}[t]
    \centering
    \includegraphics[width=\linewidth]{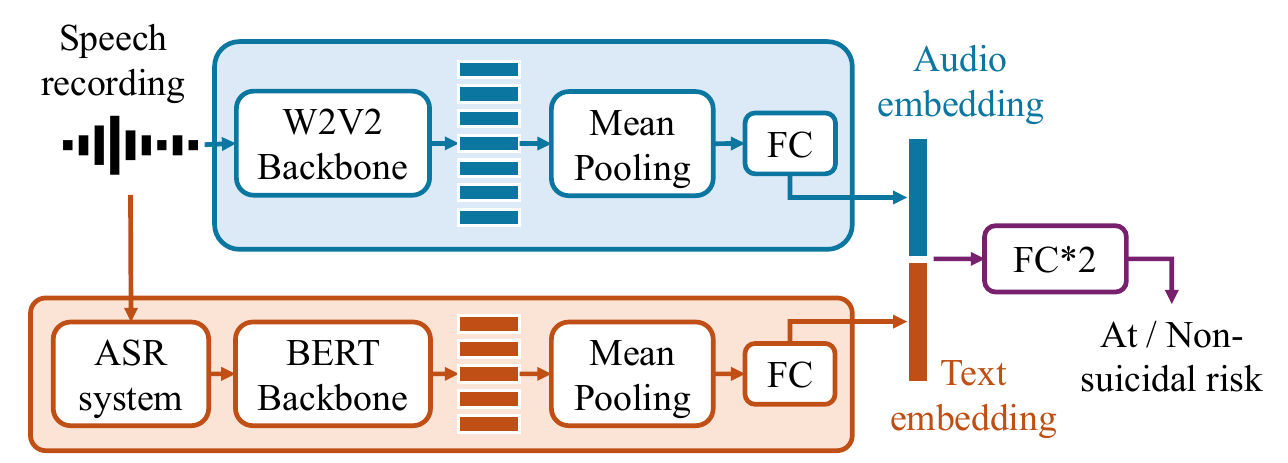}
    \caption{Illustration of the structure of the W2V2+BERT baseline. FC denotes a fully-connected layer.}
    \vskip -1ex
    \label{fig: struc}
\end{figure}

\section{Baseline Systems}
Two baseline systems were provided for the challenge: ``Baseline'' system based on hand-crafted features and traditional machine learning classifiers, and ``Baseline-bonus'' system based on foundation models and neural network classifiers.

\subsection{Baseline: eGeMAPS+SVM}
The extended Geneva Minimalistic Acoustic Parameter Set (eGeMAPS) is a standard set of acoustic features commonly used in speech analysis. It contains the fundamental frequency (F0), energy features (\textit{e.g.}, loudness), spectral features (\textit{e.g.}, formant frequencies, HNR), voice quality (\textit{e.g.}, jitter, shimmer), as well as their most common statistical functionals, for a total of 88 features per speech segment. The eGeMAPS+SVM baseline extracts the eGeMAPS feature set from the input speech segment and feeds the features into a support vector machine (SVM) for classification. Results of the three tasks were combined by voting. 
The system was implemented using the scikit-learn toolkit. The ER and ED tasks used RBF kernel with $\gamma=100$ while the PR task used linear kernel. The L2 penalty coefficient was set to 10 for all tasks. The default values were kept for all other hyperparameters.

The eGeMAPS+SVM system serves as the standard baseline, a reference point for challenge teams to compare against.

\subsection{Baseline-bonus: W2V2+BERT}
The overall structure of the W2V2+BERT baseline is shown in Figure~\ref{fig: struc}, which contains an audio branch and a text branch.
The backbone of the audio branch uses a wav2vec 2.0 (W2V2) model\footnote{Available at: \url{https://huggingface.co/facebook/wav2vec2-large-xlsr-53}}~\cite{baevski2020wav2vec}, 
which contains 24 Transformer blocks with model dimension 1024 and a total number of 317M parameters, pretrained on 56k hours of multilingual speech data. 
The backbone of the text branch uses a BERT model\footnote{Available at: \url{https://huggingface.co/google-bert/bert-base-chinese}}~\cite{devlin2018bert}, which contains 12 Transformer blocks with model dimension 768 and a total number of 110M parameters, pretrained on Chinese with a vocabulary size of 21,128. 

The training consisted of two stages. In the first stage, the W2V2 and the BERT backbones were separately finetuned for suicide risk detection. In the second stage, features extracted from the two backbones were concatenated and used to train the fused suicide risk classifier. For the PR task where participants read the same paragraph, only the W2V2 branch was used. The models were trained by cross-entropy loss. The predictions from the three tasks were combined by averaging their predictive probabilities (softmax outputs), and the final prediction was determined as the class with higher averaged probability. 

The models were implemented using PyTorch with Adam optimiser and Cosine learning rate scheduler. The learning rate and batch size used for each task and stage are listed in Table~\ref{tab: lr}. The speech recordings were chunked before feeding into the W2V2 model with a chunk size of  \SI{10}{s} and a chunk shift of \SI{6}{s}. All experiments were conducted on a single Nvidia 4090 GPU.

The W2V2+BERT system serves as the bonus baseline, an additional challenge designed for teams aiming to exceed a higher standard and earn a bonus distinction.

\begin{table}
    \centering
    \caption{Training parameters for the W2V2+BERT baseline.}
    \vskip -1ex
    \begin{tabular}{cccccc}
    \toprule
 & & \multicolumn{3}{c}{\textbf{Learning rate}} &{\textbf{Batch }}\\

          &&  \textbf{ER}&  \textbf{PR}& \textbf{ED} & \textbf{size}\\
           \midrule
          \multirow{2}{*}{Stage 1}&Finetune W2V2&  2e-5&  5e-5& 5e-5 & 8 \\
          &Finetune BERT&  5e-5&  N/A& 5e-5 & 16 \\
          \midrule
          Stage 2&Fusion &  1e-5&  1e-3& 5e-5 & 16 \\
          \bottomrule
    \end{tabular}
    \vskip -1ex
    
    \label{tab: lr}
\end{table}
\begin{table}[t]
\caption{Accuracy of the two baselines on the dev and test sets. }
\vskip -1ex
\setlength{\tabcolsep}{4pt}
    \centering
    \begin{tabular}{ccc}
    \toprule
         \textbf{Dev / Test} &  \textbf{eGeMAPs+SVM}&  \textbf{W2V2+BERT}\\
         \midrule
         Emotional Regulation& 0.52 / 0.51 &  0.56 / 0.54\\
         Paragraph Reading& 0.54 / 0.48 & 0.57 / 0.54 \\
         Expression Description& 0.58 / 0.57 & 0.54 / 0.60\\
         \midrule
         Combined & 0.53 / 0.51 & 0.56 / {0.61} \\
         \bottomrule
    \end{tabular}
    \vskip -2.5ex
    \label{tab: results}
\end{table}

\subsection{Results}
The performance is evaluated by accuracy and the results are listed in Table~\ref{tab: results}. Overall, W2V2+BERT system outperforms eGeMAPs+SVM system. Among the three tasks, the ED task achieves slightly better results. 
However, the combination of the three tasks does not necessarily lead to improvement, possibly due to the limited capability of the models for each individual task.  
Challenge teams are encouraged to explore more advanced models and sophisticated techniques to enhance suicide risk detection performance. They may choose to use recordings from a single task or combine data from different tasks.
The combination of three tasks in the W2V2+BERT system yields an accuracy of 56\% on the dev set and 61\% on the test set. For the original speech before timbre anonymisation, the accuracy is 58\% on the dev set and 60\% on the test set, which suggests that the anonymisation process preserves the information relevant to the task.

\begin{figure}[t]
    \centering
    \includegraphics[width=0.95\linewidth]{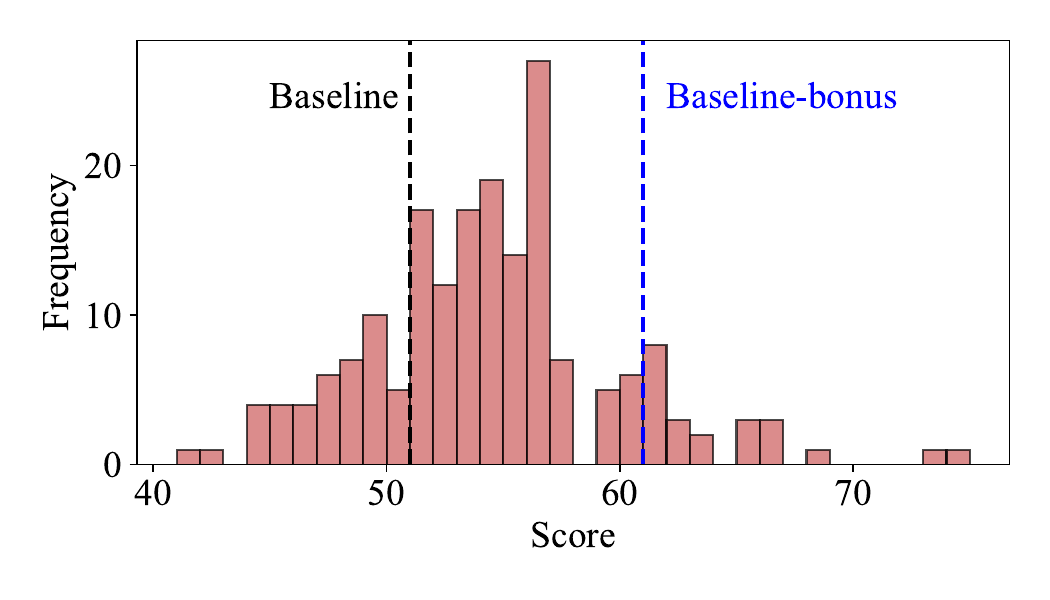}
    \caption{Distribution of scores from challenge submissions.}
    \label{fig: scores}
\end{figure}

\section{Summary of Participation}
The SW1 challenge has received 127 registrations and 186 system submissions from teams worldwide. 
The score distribution is visualised in Figure~\ref{fig: scores}. We are pleased to report that 68.8\% of the systems surpassed the Baseline system and 14 systems beat the Baseline-bonus! The top-performing system achieved an accuracy of 0.74 on the test set,  representing a 21.3\% relative improvement over the Baseline-bonus. The score distribution provides insight into the relative difficulty of the task and the effectiveness of different approaches.

\section{Conclusion}
This paper introduces the 1st SpeechWellness Challenge (SW1), the first challenge targeting at current suicide risk detection among adolescents through speech analysis. 
The dataset collected for the challenge comprises speech recordings from 600 adolescents, each performing three speech tasks: emotional regulation question, passage reading, and expression description. All recordings have undergone timbre modification to ensure anonymisation and protect the privacy of the participants. A further filtering process was also applied to ensure the effectiveness of the responses, and no sensitive information is disclosed. Apart from the standard Baseline system, we also introduced a Baseline-bonus system for the challenge, targeting teams aiming to exceed a higher benchmark and achieve a bonus distinction.
The challenge received 127 registrations, and the best-performing system achieved an accuracy of 0.74 on the test set.
The SW1 challenge seeks to highlight the importance of mental health in speech technology research, advances the development of novel algorithms to detect subtle cues in speech related to current suicide risk, and foster collaboration across the fields of speech technology, psychology, and healthcare.

\section{Limitations and Ethical Consideration}

This study’s findings are based on the MINI-KID scale, which assesses current suicide risk based on participants' immediate responses. While widely recognised as a gold standard for adolescent suicide risk assessment, MINI-KID has limitations, including reliance on self-reported data, which may not fully capture the complexity of suicidal ideation. These results should not be interpreted as predictions of future suicidal behaviour and are strictly confined to this assessment’s context.

Generalisation is a common challenge in machine learning, especially in this task, where privacy concerns and the sensitive nature of the issue limit data availability and variety. The test set labels were withheld during the challenge to ensure systems were evaluated on unseen data. However, the dataset may not fully reflect the diversity of suicidal behaviours across different populations, as factors like culture and socioeconomic status could affect how generalizable the findings are.

Automatic prediction of suicide risk remains challenging, as reflected in our results, with an inherent risk of misclassification.  This issue extends beyond suicide risk to medical tasks where errors have serious consequences. Developing such systems is still an ongoing process, requiring continuous refinement of models, improved data representation, and interdisciplinary collaboration -- precisely why we launched this challenge. Given the complexity of this task and the need for further advancements, the SW1 challenge aims to encourage innovative solutions, foster collaboration, and benchmark approaches to advance reliable, ethical suicide risk prediction models.

\bibliographystyle{IEEEtran}
\bibliography{mybib}

\end{document}